\documentclass[aps, pre, twocolumn, superscriptaddress, reprint, longbibliography]{revtex4-1}

\usepackage{amsfonts, amsmath, bm, graphicx}
\def\vec#1{{\bm{#1}}}
\def\mat#1{{\hat{\vec{#1}}}}
\def\H{{\mathcal{H}}}
\def\F{{\mathcal{F}}}
\def\t#1{{\mathrm{#1}}}

\begin{document}

\title{Controlling selection rules for magnon scattering in nanomagnets\\by spatial symmetry breaking}

\author{Arezoo Etesamirad}
\thanks{These authors contributed equally to this work}
\affiliation{Physics and Astronomy, University of California, Riverside, CA 92521, USA}

\author{Julia Kharlan}
\thanks{These authors contributed equally to this work}
\affiliation{Institute of Magnetism, Kyiv 03142, Ukraine}
\affiliation{Institute of Spintronics and Quantum Information, Faculty of Physics, Adam Mickiewicz University, Pozna\'{n}, Poland}

\author{Rodolfo Rodriguez}
\affiliation{Physics and Astronomy, University of California, Riverside, CA 92521, USA}

\author{Igor Barsukov} \email[corresponding author, e-mail: ]{igorb@ucr.edu}
\affiliation{Physics and Astronomy, University of California, Riverside, CA 92521, USA}

\author{Roman Verba} \email[corresponding author, e-mail: ]{verrv@ukr.net}
\affiliation{Institute of Magnetism, Kyiv 03142, Ukraine}

\date\today

\begin{abstract}
Nanomagnets are the building blocks of many existing and emergent spintronic technologies. Magnetization dynamics of nanomagnets is often dominated by nonlinear processes, which have been recently shown to have many surprising features and far-reaching implications for applications. Here we develop a theoretical framework uncovering the selection rules for multimagnon processes and discuss their underlying mechanisms. For its technological relevance, we focus on the degenerate three-magnon process in thin elliptical nanodisks to illustrate our findings. We parameterize the selection rules through a set of magnon interaction coefficients which we calculate using micromagnetic simulations. We postulate the selection rules and investigate how they are altered by perturbations, that break the symmetry of static magnetization configuration and spatial spin-wave profiles and that can be realized by applying off-symmetry-axis or nonuniform magnetic fields. Our work provides the phenomenological understanding into the mechanics of magnon interaction as well as the formalism for determining the interaction coefficients from simulations and experimental data. Our results serve as a guide to analyze magnon processes inherently present in spin-torque devices for boosting their performance or to engineer a specific nonlinear response of a nanomagnet used in neuromorphic or quantum magnonic application.
\end{abstract}

\maketitle


\section{Introduction}

Nonlinear magnetization dynamics is an exciting field of fundamental physics which bears tremendous potential for applications in information technologies and beyond \cite{Kosevich_Physica1981, Cottam_Book1994, Wigen_Book1994, Gurevich_Book1996, Bertotti_Book,Kosevich_PhysRep1990,CherkasskiiPRB2022,demidov_magnetic_2012,YiLiHybrid,duan2014nanowire}. In contrast to many other physical systems, nonlinearity is inherent to magnetic systems and of topological origin \cite{Prokopenko_UJP2019} -- the phase space for magnetization vector motion is not a plane, but a sphere $|\vec M| = M_s$. This results in nonlinearity although most magnetic interactions (exchange, dipolar, uniaxial anisotropy, Dzyaloshinskii-Moriya interaction) are quadratic functions of the magnetization $\vec M$. Nonlinear processes can thus be observed at relatively low excitation levels and exploited in applications --  in particular in microwave electronics \cite{Gurevich_Book1996, How_Book2005, Geiler_IEEEJM2021}, analog and digital signal processing \cite{Chumak_NatPhys2015, Pirro_NRM2021, Chumak_Roadmap2022}, non-Boolean computing such as magnetic neuromorphics \cite{Torrejon_Nat2017, Grollier_NatEl2020} and quantum information systems \cite{Tabuchi405,Lachance-Quirion425,LuqiaoMagnonPhonon}.

At up to moderately high excitation levels, nonlinear dynamics is often treated as interaction of linear spin wave modes or, within the quantum picture, as scattering of magnons (quanta of spin waves) \cite{Schlomann_PR1959, White_PR1963, Lvov_Book1994, Safonov_Book}. For instance, two-magnon processes can be the dominant contribution to damping in thin films \cite{BarsukovIEEE2010}, three-magnon processes can lead to parametric magnon excitation \cite{Chen_NanoLett2017}, and four-magnon processes \cite{HammelPRAppl2022} are responsible for magnon thermalization and condensation \cite{safranski2017spin,demokritov2006bose}.

Magnon processes have been extensively studied in bulk samples and thin ferromagnetic films (see, e.g. Refs.\,\cite{Lvov_Book1994, Cottam_Book1994, Wigen_Book1994, Boardman_PRB1988, Livesey_PRB2007} and references therein). However, the obtained results cannot be directly transferred to the case of micron- and nanoscale finite-size magnetic structures. First, under strong geometric confinement, the spin wave spectrum is discrete. The magnon processes are resonant and occur principally within a well-defined parameter sub-space (e.g. at particular external fields) \cite{Boone_PRL2009, Melkov_MagLet2013, Slobodianiuk_MagLett2019, Kobljanskyj_SR2012}. The discreteness of the magnon spectrum in micro-/nano-magnets can lead to qualitatively different behavior of magnon processes as compared to geometrically extended systems \cite{Barsukov_SciAdv2019}. Second, the spin-wave eigenmodes can no longer be considered plane-waves like in bulk samples and films. Their spatial profile as well as the static magnetization configuration depend on the shape of the magnet. Consequently, each micro-/nano-magnet possesses an individual static magnetization configuration and spin wave profiles, thus subjecting magnon processes to a set of specific selection rules \cite{Schultheiss_PRL2019, Camley_PRB2014}. Understanding, predicting, and controlling theses selection rules is instrumental for designing functional spintronic applications.

In this work, we parameterize the selection rules through a set of magnon coupling coefficients for magnon scattering processes. We systematically study the effect of the symmetry of the static magnetization and spin wave profiles on the magnon coupling coefficients. Our results provide a comprehensive guide to understanding and engineering nonlinear phenomena in micro-/nano-magnets.

While the conceptual approach of our study can be extended to a variety of magnon processes, here, we mainly focus on the degenerate three-magnon confluence process, in which two magnons of one kind fuse into a single magnon of another kind. In nanomagnets, three-magnon processes show a substantial effect on magnetization dynamics even at low excitation levels \cite{Boone_PRL2009, Schultheiss_PRL2009, Nguyen_PRB2011, Barsukov_SciAdv2019}, often constitute the main dissipation channel of primary magnons, and can be used for enhancing functionality of spintronic applications e.g. through frequency doubling \cite{Demidov_PRB2011, Gruszecki_APL2021}. Moreover, three-magnon processes have been recently shown to invert a nanomagnet's response to spin-torque \cite{Barsukov_SciAdv2019}, thus having far-reaching implications for spin-torque devices and potentially on magnetic neural networks.

In our recent work \cite{Etesamirad_ACSAMI2021}, we experimentally demonstrated how a magnon coupling coefficient can be manipulated by altering the symmetry of spin wave profiles via application of magnetic field with nanoscale nonuniformity. On the basis of such proof-of-concept, we systematically investigate avenues to controlling magnon interaction and seek to provide a manual for engineering nonlinearity in nanomagnets. As a sample platform for our study, we consider thin elliptic ferromagnetic nanodisks in single-domain magnetization state (particular attention is paid to the quasiuniform state). Nonetheless, the results of our study are directly applicable to other shapes of nanomagnets that possess mirror symmetry respective to two perpendicular in-plane axes, e.g., rectangular, stadium-shaped, etc. At the same time, the conceptual inferences, made in this paper, are expected to be applicable universally to any thin nanomagnet.

The article is organized as follows: In Section~\ref{s:theory} we describe the vectorial Hamiltonian formalism for nonlinear spin-wave dynamics, which lays the basis for calculation of the three-magnon coupling coefficients. Section~\ref{s:methods} introduces micromagnetic simulation methods of our study. In Section~\ref{s:sym}, we investigate three-magnon process selection rules in a systems with intrinsic, unperturbed symmetry of magnetic configuration and spin-wave profiles. In Section~\ref{s:sym_break}, we investigate the effect of symmetry-breaking perturbation fields on the magnon coupling coefficients and  discuss the routes to engineering magnon coupling in experiments. Finally, conclusion are given in Sec.~\ref{s:summary}.      


\section{Theoretical basis}\label{s:theory}

\subsection{Vectorial Hamiltonian formalism}

We use the recently developed vectorial Hamiltonian formalism \cite{Tyberkevych_ArXiv} which allows one to easily account for spatially nonuniform configuration of static magnetization and complicated spatial spin-wave profiles. The standard scalar Hamiltonian formalism, usually used for spatially-uniform ground state \cite{Lvov_Book1994, Safonov_Book}, can be generalized to non-uniform case, e.g. a domain wall \cite{Abyzov_JETP1979}. However, it is typically used with analytically-defined magnetization states and magnon modes. Here, we will implement static magnetization configurations and spin-wave profiles obtained from micromagnetic simulations and resort to the vector Hamiltonian formalism, which has been successfully employed for studies of nonlinear frequency shift of edge modes \cite{Tyberkevych_ArXiv}, three-magnon splitting in vortex-state magnetic dots \cite{Verba_PRB2021} and nanotubes \cite{Korber_PRB2022}.   

The dynamics of a constant-amplitude three-dimensional magnetization vector on a unit sphere $|\vec M(\vec r, t)|/M_s = 1$ is mapped to the dynamics of a two-dimensional vector of dynamic magnetization on a plane disk. This mapping is analogous to the Lambert azimuthal equal-area projection \cite{Snyder_1987}:
\begin{equation}\label{e:Lambert}
      \frac{\vec M(\vec r, t)}{M_s} = \left(1-\frac{|\vec s(\vec r, t)|^2}{2} \right) \vec \mu(\vec r) + \sqrt{1-\frac{|\vec s(\vec r, t)|^2}{4}} \vec s (\vec r,t) \ .
\end{equation}
Here $\vec\mu (\vec r) = \vec M_0(\vec r)/M_s$ is the spatial configuration of the normalized static magnetization, $M_s$ is the saturation magnetization, and $\vec s(\vec r, t)$  is the normalized dynamic magnetization, which is perpendicular to the static one, $\vec s \bot \vec \mu$. The dynamic magnetization can be expanded in a series of linear spin-wave eigenmodes $\vec s_\nu$ of the system:
  \begin{equation}\label{e:s-exp}
   \vec s(\vec r, t) = \sum\limits_\nu \left(c_\nu(t) \vec s_\nu (\vec r) + \t{c.c.} \right) \,,
  \end{equation}
where $c_\nu$ are complex amplitudes of the spin-wave eigenmodes. To arrive at the equations of motion for spin-wave eigenmodes in a standard Hamiltonian form, spatial profiles of linear eigenmodes are normalized:
  \begin{equation}\label{e:norm}
    \frac{i}{V} \int \vec s_\nu^* \cdot \vec\mu \times \vec s_\nu \,d^3\vec r = 1 \,,
  \end{equation}
where the integration goes over all the sample volume $V$. Note, that, following P.~Krivosik and C.~Patton \cite{Krivosik_PRB2010}, here we use a normalized spin-wave Hamiltonian $\H = \gamma E/(M_s V)$ which is measured in the units of frequency, where $E$ is the total magnetic energy. This approach is convenient for classical magnetic systems (for normalizations of quantum system, see e.g. \cite{Kusminskiy2021, Romero2020}), as the variable $\vec s$ has a clear sense of dimensionless (normalized per $M_s$) dynamic magnetization, and all the coefficients of Hamiltonian expansion, including three-wave coefficients $V_{\nu\eta,\zeta}$, are of the same frequency units having the sense of effective interaction frequencies \cite{Krivosik_PRB2010}. Within this approach, normalization Eq.~\eqref{e:norm} ensures that quadratic part of the Hamiltonian assumes the standard form in terms of the spin-wave amplitudes:  $\H^{(2)} = \sum_\nu |c_\nu|^2\omega_\nu$ \cite{Tyberkevych_ArXiv}.

The three-wave term of the spin-wave Hamiltonian can be written as:
  \begin{equation}\label{e:H3-gen}
   \H^{(3)} = -\frac{\omega_M}{2V} \int (|\vec s|^2 \vec \mu) \cdot \mat N \cdot \vec s \,d^3\vec r \,,
  \end{equation}
where $\omega_M = \gamma \mu_0 M_s$, $\gamma$ is the modulus of gyromagnetic ratio and $\mat N$ is a tensor operator describing magnetic self-interactions.

\textbf{The interaction operator.} The operator consists of exchange, dipolar, anisotropy, and other contributions: $\mat N = \mat N^\t{(ex)} + \mat N^\t{(dip)} + \mat N^\t{(an)} + \dots$

The exchange operator is given by 
\begin{equation}\label{e:N_ex}
 \mat N^\t{(ex)} = -\lambda_\t{ex}^2 \mat I \nabla^2 \,,
\end{equation}
where $\lambda_\t{ex}$ is the exchange length and $\mat I$ is the identity matrix. 

The uniaxial anisotropy operator is
\begin{equation}\label{e:N_an}
 \mat N^\t{(an)} = -\frac{B_\t{an}}{\mu_0 M_s} \vec e_{z'} \otimes \vec e_{z'} \,,
\end{equation}
where $B_\t{an} = 2 K_u /M_s$ is the anisotropy field, $K_u$ is the anisotropy constant, $\vec e_{z'}$ is the unit vector of the anisotropy axis, and $\otimes$ denotes dyadic product of vectors. If the coordinate system is chosen such that the anisotropy axis coincides with a coordinate axis (e.g., $z$-axis), the anisotropy operator has only one nonzero component (e.g., $N_{zz}$). 

The operator describing the magnetodipolar interaction is expressed through the magnetostatic Green function $\mat G$:
  \begin{equation}\label{e:Ndip-s}
   \mat N^\t{(dip)} \cdot \vec s = \int \mat G (\vec r, \vec r') \cdot \vec s(\vec r') \,d^3 \vec r'.
  \end{equation}
In thin disks, which we consider here, magnetization along the thickness coordinate ($z$) can be assumed uniform. In this case, integration over the volume $V$ in Eqs.~(\ref{e:H3-gen}, \ref{e:Ndip-s}) is changed to the integration over the sample area $S$, and the magnetostatic Green function can therefore be conveniently expressed via its Fourier-transform
  \begin{equation}\label{e:G-Nk}
   \mat G (\vec r, \vec r') = \frac1{4\pi^2} \int \mat N_{\vec k}^\t{(dip)} e^{i \vec k \cdot (\vec r - \vec r')} \,d^2\vec k \,, 
  \end{equation}
where $\vec k = k_x \vec e_x + k_y \vec e_y$ is a two-dimensional in-plane wave vector and $\mat N_{\vec k}^\t{(dip)}$ is defined in Cartesian components as  \cite{Guslienko_JMMM2011} 
  \begin{equation}\label{e:Nk}
    \mat N_\vec k = \left( \begin{array}{ccc}
                            k_x^2 f(kh) / k^2 & k_x k_y f(kh) / k^2 & 0 \\
                            k_x k_y f(kh) / k^2 & k_y^2 f(kh) / k^2 & 0 \\
                            0 & & 1-f(kh)
                           \end{array}
   \right) \ . 
  \end{equation}
Here, $f(x) = 1 - (1-e^{-|x|})/|x|$ is the so-called ``thin film function'' with the sample thickness $h$. 

\textbf{Interaction coefficients.} Using the eigenmode expansion \eqref{e:s-exp}, Eq.~\eqref{e:H3-gen} is transformed to the standard form of spin-wave mode interaction:
  \begin{equation}
       \H^{(3)} = \frac13 \sum\limits_{\nu\eta\zeta} \left(U_{\nu\eta\zeta}c_\nu c_\eta c_\zeta + \t{c.c.} \right) + \sum\limits_{\nu\eta\zeta} \left(V_{\nu\eta,\zeta}c_\nu c_\eta c_\zeta^* + \t{c.c.} \right)
  \end{equation}
The first sum here describes creation of three magnons and inverse process of magnon annihilation, which can be resonant only in active media (so called ``explosive instability'' \cite{Lvov_Book1994}), while the second sum corresponds to three-magnon confluence and splitting.

Here, we consider degenerate three-magnon confluence process where two magnons of the mode ``$\nu$'' fuse into one magnon of the mode ``$\eta$'' (in short notation, $(\nu+\nu) \to \eta$). This process is described by the term $V_{\nu\nu,\eta}~ c_{\nu} c_{\nu} c_{\eta}^*$, and the corresponding three-magnon (coupling) coefficient is given by:
  \begin{equation}\label{e:V112}
  \begin{split}
   V_{\nu\nu,\eta} &= -\frac{\omega_M}{2V} \int \left(2(\vec s_{\nu} \cdot \vec s_{\eta}^*) \vec\mu \cdot \mat N \cdot \vec s_{\nu} \right. \\
   &+ \left. ({\vec s_{\nu}} \cdot \vec s_{\nu}) \vec\mu \cdot \mat N \cdot \vec s_{\eta}^*\right)d^3\vec r \ .
  \end{split}
  \end{equation}
This equation can be used for explicit calculation of the three-magnon confluence coefficients $V_{\nu\nu,\eta}$. As input, it requires the static magnetization configuration and spin-wave modes profiles.  They can be extracted from micromagnetic simulations, other numerical methods, or analytical approximation (in simple structures or as a zero approximation). In our effectively two-dimensional geometry, the dipolar contribution to the magnon coupling coefficient \eqref{e:V112} (term with $\mat N^\t{(dip)}$) can conveniently expressed via Fourier-images as
  \begin{equation}\label{e:V112-dip}
      \begin{split}
	V_{\nu\nu,\eta}^\t{(dip)} &= -\frac{\omega_M}{8\pi^2 S} \int \left(2\F_\vec k \left[(\vec s_\nu \cdot \vec s_\eta^*) \vec\mu \right] \cdot \mat N_{\vec k}^\t{(dip)} \cdot \F_{-\vec k} \left[\vec s_\nu\right] \right. \\
	&+ \left. \F_\vec k \left[({\vec s_\nu} \cdot \vec s_\nu) \vec\mu \right] \cdot \mat N_{\vec k}^\t{(dip)} \cdot \F_{-\vec k} \left[\vec s_\eta^*\right] \right)d^2\vec k \,,
      \end{split}   
  \end{equation}
where 
  \begin{equation}
    \F_\vec k [\vec s] = \int \vec s(\vec r) e^{i \vec k \cdot \vec r} d^2 \vec r
  \end{equation}
is the two-dimensional Fourier transform.

Equation \eqref{e:V112-dip} can be used to phenomenologically assess how the symmetry of the static magnetization configuration and spin wave profiles affect the three-magnon coefficient. Utilization of Fourier-image in Eq.~\eqref{e:V112-dip} also reduces computational  complexity of the evaluation of magnetodipolar contribution to three-magnon coefficients; other contributions -- exchange and anisotropy -- are directly calculated in coordinate space according to Eq.~\eqref{e:V112}. 
  
\subsection{Dynamics of interacting modes}
  
Equations of motion for spin-wave amplitudes are derived from the Hamilton formalism as $dc_\nu / dt = -i \partial \H/\partial c_\nu^*$. We consider the mode $\nu$ excited by an external drive with amplitude $f_e$ and frequency $\omega_e$. The mode $\eta$ is excited via the three-magnon process. The equations of motions read
\begin{equation}\label{e:dc-nu-sys}
 \begin{split}
  \frac{dc_{\nu}}{dt} &+ i \omega_{\nu} c_{\nu} + \Gamma_{\nu} c_{\nu} = -2i V_{\nu\nu,\eta}^* c_{\nu}^* c_{\eta} + f_e e^{-i\omega_e t} \,, \\
  \frac{dc_{\eta}}{dt} &+ i \omega_{\eta} c_{\eta} + \Gamma_{\eta} c_{\eta} = -i V_{\nu\nu,\eta} c_{\nu}^2 \ .
 \end{split}
\end{equation}
Here, $\omega_\nu$ are the mode eigenfrequencies and $\Gamma_\nu$ are the intrinsic damping rates of the spin-wave modes. A detailed analysis of the three-magnon dynamics including its interaction with spin-torque can be found in Ref.\,\cite{Barsukov_SciAdv2019}.

It is worth pointing out two useful relations. If the excitation levels are not too high, the spin-wave modes $\nu$ and $\eta$ can be assumed to oscillate at single and double excitation frequency, $c_{\nu} \sim e^{-i\omega_e t}$ and $c_{\eta} \sim e^{-2i\omega_e t}$, while higher harmonics can be neglected. Then, one can write the ratio of spin-wave amplitudes as
  \begin{equation}\label{e:c2-c1}
    \frac{c_{\eta}}{c_{\nu}^2} = \frac{-i V_{\nu\nu,\eta}}{i(\omega_{\eta} - 2\omega_e) + \Gamma_{\eta}} \ .
  \end{equation}
Another consequence of the three-magnon process is the enhancement of the effective damping of mode $\nu$ by the value $ 2|V_{\nu\nu,\eta}|^2 |c_\nu|^2 \Gamma_\eta/(\Gamma_\eta^2 + (\omega_\eta - 2\omega_e)^2)$, which is a directly detectable experimental evidence of three-magnon confluence \cite{Barsukov_SciAdv2019} (general case of an arbitrary excitation level is considered in details in Ref.\,\cite{Barsukov_SciAdv2019}).

\textbf{Calculation of interaction coefficients.} We thus have found two approaches to quantitatively determining magnon interaction coefficients $V_{\nu\nu,\eta}$. (i)~In what follows, we will use Eq.~\eqref{e:V112} which requires static magnetization configuration $\vec{\mu}$ and spin-wave profiles $\vec{s}_{\nu}$ as input. These input parameters will typically be obtained from micromagnetic simulations. (ii)~For validation purposes in some instances, we shall also resort to Eq.~\eqref{e:c2-c1} which requires mode amplitudes $c_{\nu}$ -- we will extract these from micromagnetic simulations as well.

Potentially, the input parameters for the both said equations could be obtained from \textit{experiment} \cite{MeckenstockJAP2006, STFMR, hickenMapping}, allowing to directly determine the interaction coefficients. For instance, the mode amplitudes for Eq.~\eqref{e:c2-c1} can often be determined directly, as can the damping variations \cite{STFMR}. Mapping static magnetization configuration and spin wave profiles is experimentally more challenging \cite{MeckenstockJAP2006, STFMR, hickenMapping}, which makes Eq.~\eqref{e:V112} a valuable but mostly theoretical tool.


\section{Micromagnetic simulations}\label{s:methods}

\begin{figure}
 \includegraphics[width=\columnwidth]{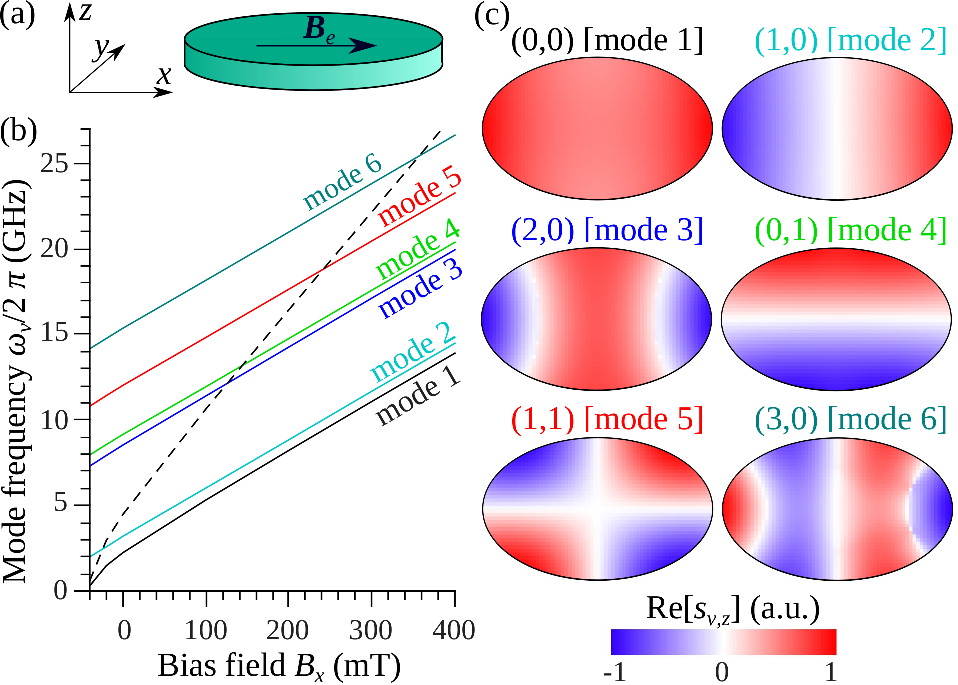}
 \caption{(a) The sample model is a thin elliptical disk in a bias magnetic field $\vec{B}_e$. (b) Bias field dependence of the first six spin-wave modes' eigenfrequencies for $\vec{B}_e \parallel \vec{e}_x$. The dashed line shows the double frequency of the lowest mode (quasiuniform, $\nu=1$). (c) Spin-wave profiles of the spin-wave modes at $B_e = 10\,$mT.}\label{f:1}
\end{figure}

Figure \ref{f:1}(a) shows the sample model used in our micromagnetic simulations using MuMax$3$ software \cite{Vansteenkiste_AIPAdv2014}. The disk is representative of samples with mirror symmetry with respect to its axes ($x$ and $y$ coordinate axes), which is reflected in the symmetry of the static magnetization configuration and spin-wave modes.

The sample parameters were chosen to mimic the CoFeB-based nanodevices used in various experimental studies \cite{BarsukovAPL2014,BarsukovAPL2015}. Lateral dimensions of the disk are $64\,\t{nm} \times 40 \,\t{nm}$, the thickness is $h = 1.5\,$nm. Saturation magnetization is $M_s =1.6\times10^6\,\t{A}/\t{m}$, exchange stiffness is $A_\t{ex} = 2 \times 10^{-11}\,\t{J}/\t{m}$. Surface perpendicular anisotropy is $K_s = 1.8\,\t{mJ}/\t{m}^2$ (accounted for as the effective volume uniaxial anisotropy $K_u = 1.2\times 10^6\,\t{J}/\t{m}^3$ in simulations), leading--with the demagnetization--to a total easy-plane magnetic anisotropy of the sample. The Gilbert damping constant is $\alpha_G = 0.007$. A cell size of $1\times 1 \times 1.5 \,\t{nm}^3$ was used.
  
Spin-wave mode spectra were simulated by time-domain Fourier-transform of time traces of the magnetization vector. Typically, magnetization was excited by a short field pulse $\vec b = b_y(\vec{r},t) \vec e_y$. To excite spatially nonuniform modes, excitation field was applied in one quadrant of the disk.

To obtain spin-wave modes profiles, we apply single-frequency excitation field at the eigenfrequency of the mode, $\vec b = b_y(\vec r) \cos[\omega_\nu t] \vec e_y$ and perform simulations until stationary oscillation amplitude is reached. Complex-valued spin-wave profiles are defined by 
  \begin{equation}\label{e:s-nu-sim}
   \vec s_\nu(\vec r) \sim \frac1{T_\nu} \int\limits_t^{t+T_\nu} (\vec m(\vec r,t) - \vec \mu(\vec r)) e^{i\omega_\nu t} dt
  \end{equation}
where $\vec m(\vec r,t)$ is instant real magnetization distribution in simulations. The integral over oscillation period $T_\nu$ is substituted by a sum in the evaluation of the simulations. The profiles were subsequently normalized according to Eq.~\eqref{e:norm}. For visualizing the spin-wave profiles in figures below, we plot the real part of their $z$-component.

Note, that if two spin-wave modes are adjacent in frequency, microwave field at the frequency $\omega = \omega_\nu$ excites not only the $\nu$-th mode, but also can excite neighboring modes with smaller, but still finite amplitudes. Then, processing of the simulation results according to Eq.~\eqref{e:s-nu-sim} gives an admixture of true spin-wave modes, $\vec s_\nu \to \vec s_\nu + \sum_{\eta \neq \nu} \varsigma_{\nu,\eta} \vec s_\eta$ with $\varsigma_{\nu,\eta} \sim \Gamma_\eta/|\omega_\nu - \omega_\eta + i\Gamma_\eta| < 1$. As shown below, three-magnon interaction coefficients are very sensitive to the mode symmetry, and such an admixture may result in incorrect calculation results. To avoid these artifacts, the symmetry of the excitation field was adjusted to be mode-specific, so that it excites given mode, but not neighboring ones. To achieve this, we used nonuniform excitation fields that have strong preferential excitation for one mode but not for another. Such fields can be selected based on the symmetry of the magnon mode (e.g. see \cite{Kruglyak2017}). In our case, excitation fields that are simply localized within a sub-area of the disk can be chosen. For instance, a uniform drive field was used for the $(0,0)$ mode, while the field was localized to the the upper half of the disk for $(0,1)$ mode. Alternative approach could be a reduction of the Gilbert damping, which, however, would result in an increase of simulation time.

\textbf{Spin-wave modes.} The first six spin-wave modes of the disk are shown in Fig.~\ref{f:1}(b,c). As the disk size is below the edge-to-bulk mode crossover \cite{Carlotti_APR2019}, there are no edge modes and the lowest mode is the quasiuniform $(0,0)$ mode. Higher modes are backward-volume-like modes $(n_x, 0)$, Damon-Eshbach-like modes $(0,n_y)$, and mixed modes \cite{Giovannini_PRB2004} ($n$ is the mode index, i.e. number of nodes along the $x$ or $y$ directions). When external magnetic field $\vec{B}_e$ is aligned to the long axis of the disk, $\vec B_e = B_x \vec e_x$, the modes posses strictly symmetric (even mode index $n_\alpha$) or strictly antisymmetric  (odd $n_\alpha$) behavior with respect to each ($\alpha = x, y$) axis.  The shown modes cover all possible symmetries: (S,S), (S,A), (A,S), (A,A), where ``S'' and ``A'' mean symmetric and antisymmetric, respectively.

In what follows, we mostly study degenerate magnon processes where two magnons of the lowest mode $\nu=1$ confluence into a magnon of mode $\eta>1$. The variety of modes depicted in Fig.~\ref{f:1}(c) allows us to investigate processes of various symmetry mixes. The processes involving the confluence of the lowest mode are of particular importance for spintronics applications. Nonetheless, the conclusions of our study are directly applicable to a degenerate three-magnon process involving any combination of the spin-wave modes. We shall also touch upon more general non-degenerate three-magnon processes.


\section{Three-magnon confluence without perturbations}\label{s:sym}

\subsection{Uniform in-plane magnetization configuration}

In this section, we consider the case when magnetic field is applied along the major axis of the elliptical disk, $\vec B_e = B_x \vec e_x$,  and preserves the symmetry of magnetization configuration and spin-wave modes. It is convenient to start from an idealized case of uniform magnetization, $\vec \mu = \vec e_x$. Since dynamic magnetization is $\vec s = (0,s_y,s_z) \bot \vec \mu$ and the operators of exchange interaction and uniaxial anisotropy are \emph{diagonal} (see Eqs.~(\ref{e:N_ex},\ref{e:N_an})), these interactions do not contribute to the three-magnon scattering. In fact, the exchange interaction does not contribute to the three-magnon scattering for \emph{any} uniform magnetization configuration.

For the magnetodipolar contribution, we inspect the second term in Eq.~\eqref{e:V112-dip}. The vector-function $(\vec s_{\nu} \cdot \vec s_{\nu}) \vec \mu$ has only $x$-component and is symmetric for any symmetry of mode $\vec s_{\nu}$. Its Fourier image is an even function of both $k_x$ and $k_y$. Since $s_{\eta,x} = 0$, only the $N_{\vec k, xy}^\t{(dip)}$ component is relevant -- it is an \emph{odd} function of $k_x$ and $k_y$. Thus, the integration in Eq.~\eqref{e:V112-dip} gives a nonzero value only if $\F_{\vec k}[\vec s_{\eta}^*]$ is an odd function of both $k_x$ and $k_y$. This is only possible if $\vec s_{\eta}$ is a fully antisymmetric mode, i.e. antisymmetric with respect to both $x$ and $y$ axes.

The first term in Eq.~\eqref{e:V112-dip} possesses the same features. If mode $\vec s_{\eta}$ is symmetric respective to the both $x$ and $y$ axes, then both  Fourier-images $\F_\vec k \left[(\vec s_{\nu} \cdot \vec s_{\eta}^*) \vec\mu \right]$ and  $\F_{-\vec k} \left[\vec s_{\nu}\right]$ posses the same symmetry -- they both are either even or odd functions of $k_{x,y}$, depending on the symmetry of the mode $\vec s_{\nu}$. The resulting integral with the odd function $N_{\vec k,xy}$ is zero. 

We conclude that in the idealized case of a uniform symmetric sample, three-magnon confluence is possible only into fully antisymmetric modes, e.g. into mode~5 in Fig.~\ref{f:1}(c). This rule is valid independently of the symmetry of the primary mode $\nu$. Exceptions are edge modes, present in sufficiently large samples, which -- strictly speaking -- are neither symmetric nor antisymmetric. Confluence of an edge mode into another mode is always allowed but, since edge and volume modes spatially overlap to little extent, their nonlinear interaction is weak.

\subsection{Nonuniform symmetric in-plane magnetization configuration}

The above conclusions can be generalized to a spatially nonuniform but \emph{symmetric} in-plane magnetic state. As an example, we consider the so-called ``leaf state'' in which magnetization near the edges is parallel to the edge. We represent $\vec\mu  = (\cos\varphi_M, \sin\varphi_M, 0)$ and $\vec s_\nu = (-s_{\nu,\t{ip}}\sin\varphi_M, s_{\nu,\t{ip}}\cos\varphi_M, s_{\nu, z})$, where $\varphi_M(x,y) = -\varphi_M (-x,y) = -\varphi_M(x,-y)$ is an antisymmetric function for the leaf state, and $s_{\nu, \t{ip}}$ is an in-plane dynamic magnetization component of $\nu$-th mode.

The second term under the integral in Eq.~\eqref{e:V112} is expanded as
  \begin{equation}\label{e:V-rotated}
   \begin{split}
    V_{\nu\nu,\eta} &\sim -\left(\vec s_{\nu} \cdot \vec s_{\nu} \right) \cos\varphi_M N_{xx} \sin\varphi_M s_{\eta,\t{ip}}^*\\
    &+ \left(\vec s_{\nu} \cdot \vec s_{\nu} \right) \sin\varphi_M N_{yy} \cos\varphi_M s_{\eta, \t{ip}}^* \\
    &+ \left(\vec s_{\nu} \cdot \vec s_{\nu} \right) \cos\varphi_M N_{xy} \cos\varphi_M s_{\eta,\t{ip}}^* \\
    &- \left(\vec s_{\nu} \cdot \vec s_{\nu} \right) \sin\varphi_M N_{yx} \sin\varphi_M s_{\eta,\t{ip}}^* \ .
   \end{split}
  \end{equation}
The terms with diagonal components $N_{xx}$ and $N_{yy}$  contain the antisymmetric function $\sin\varphi_M$ and thus give a nonzero contribution only if $s_{\eta,\t{ip}}$ is antisymmetric. Terms with off-diagonal components contain the symmetric function   $\cos^2\varphi_M$ or $\sin^2\varphi_M$, thus yielding the same selection rules as for the uniform magnetization configuration discussed above. Analyzing the first term in Eq.~\eqref{e:V112} yields the same selection rules.

For the case of symmetric nonuniform magnetization configuration, it also should be noted that the exchange operator does not contribute to the three-magnon interaction since it is diagonal and symmetric (its Fourier-representation is proportional to $|\vec k|^2$).  The uniaxial anisotropy operator also does not contribute if the anisotropy axis is aligned to the dot symmetry axes (i.e. $x$, $y$ or $z$ axis).

We conclude that, in a nonuniform but symmetric magnetization configuration, three-magnon confluence is possible into a \emph{fully antisymmetric} mode only. Parenthetically should be mentioned that a similar behavior has been reported in Ref.\,\cite{Demidov_PRB2011} for propagating spin-waves in magnetic stripes. Also, in the inversion of the confluence process, only a fully antisymmetric can undergo a degenerate three-magnon \textit{splitting}. In fact, suppression of the three-magnon splitting process was found in other magnetic structures with high symmetry: in vortex-state circular magnetic dots \cite{Schultheiss_PRL2019, Verba_PRB2021}, radial modes can undergo only nondegenerate three-magnon splitting, i.e. into a pair of different modes; in vortex-state magnetic nanotubes, the same restriction applies to the modes with zero wave vector along the nanotube axis \cite{Korber_PRB2022}.

\subsection{Nondegenerate magnon processes}

The above analysis can be extended to the interaction of three disparate modes. It shows that three-magnon interaction of two symmetric modes with an antisymmetric one is allowed. Interaction of three antisymmetric modes is allowed as well, while interaction of two antisymmetric modes with one symmetric and interaction of three symmetric modes are prohibited. We can formulate the following \emph{selection rule}: both sums $\sum_\nu n_{\nu,x}$ and $\sum_\nu n_{\nu,y}$ over the indices of three interacting modes should be \emph{odd} numbers (note that an odd $n$ corresponds to an antisymmetric mode profile in the used notations). 

\subsection{Other uniform magnetization states}

All above-formulated rules for degenerate and nondegenerate processes also apply for the case when the disk is magnetized along its minor axis, $\vec\mu = \vec e_y$. For the perpendicular magnetization, $\vec\mu = \vec e_z$, the situation is different. Since off-diagonal components $N_{xz} = N_{yz} = 0$ will vanish for all operators, three-magnon interaction is completely prohibited in thin disks, strips, and films with uniform  perpendicular magnetization configuration. Only atypical anisotropy \cite{OrtizPRAppl2021, BarsukovAPL2014} with anisotropy axis which is not parallel nor perpendicular to the $z$ axis could allow for it.

It should be noted that three-magnon confluence has been experimentally observed in perpendicular nanodisks incorporated in the magnetic tunnel junctions \cite{Barsukov_SciAdv2019}, suggesting that some sample systems may substantially deviate from the idealized case discussed so far (see discussion in Sec.~\ref{ss:sym_break_summary}).

\subsection{Simulations}

We carried out a series of micromagnetic simulations to evaluate static magnetization configuration and spin-wave profiles and calculated the three-magnon interaction coefficients according to Eq.~\ref{e:V112}. The results are summarized in Table~\ref{t:1}. Confluence of two magnons of mode 1 into a magnon of the  mode 3, 4, and 6 is attributed with a vanishing (below the accuracy of our calculations) coefficient. In contrast, the process $1+1 \to 5$, i.e. confluence into a fully antisymmetric mode, is characterized by a large three-magnon coefficient $V_{11,5}$. All these features are in full agreement with the above-mentioned theoretical predictions. Small but finite three-magnon coefficient was found for the process $1+1 \to 2$ -- this result is unexpected. Further analysis shows that this process takes place at a low negative magnetic field (Fig.~\ref{f:1}(b)) associated with a strongly nonuniform magnetization configuration, which partially breaks the symmetry restrictions.  

\begin{table}
\begin{center}
\begin{tabular}{|c|c|c|}
 \hline
   Process  & Mode symmetry & $V_{11,\eta}/2\pi\,$(GHz) \\ \hline
   $1+1\to 2$ & (A,S) & $4\times 10^{-3}$ \\ \hline
   $1+1\to 3$ & (S,S) & $ < 10^{-4}$ \\ \hline
   $1+1\to 4$ & (S,A) & $ < 10^{-4}$ \\ \hline
   $1+1\to 5$ & (A,A) & $1.31$ \\ \hline
   $1+1\to 6$ & (A,S) & $ < 10^{-4}$ \\ \hline
\end{tabular}
\end{center}
\caption{Interaction coefficient for confluence processes $1+1 \to \eta$, with bias field applied along the long disk axis. Mode symmetry (S,A) describes a mode profile that is symmetric along $x$ and antisymmetric along $y$ axis, etc. }\label{t:1}
\end{table}

To validate our results, we simulated magnon confluence dynamics directly by exciting mode 1 with a  microwave field $b_z = 1\,$mT. The excitation field is spatially uniform; it thus cannot excite modes 2, 4, 5, and 6, while its coupling to mode 3 is weak. The drive  frequency is varied with the external field to coincide with the eigenfrequency of mode 1. We extracted the stationary amplitudes of the first and the second harmonics of magnetization oscillations.

\begin{figure}
 \includegraphics[width=\columnwidth]{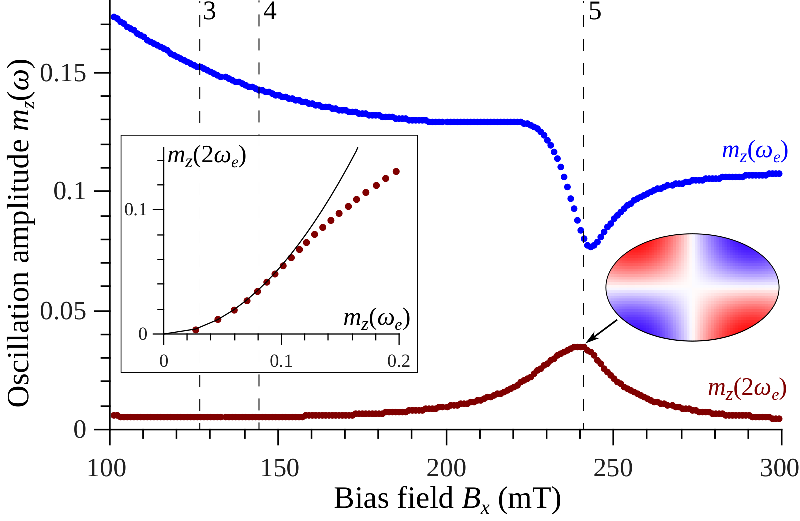}
 \caption{Amplitudes of the first and second harmonics of magnetization oscillations $m_z(t)$ excited by a uniform microwave field $b_{z} = 1\,$mT at the eigenfrequency of mode~1. The bias field is applied parallel to the long axis of the disk. Vertical dashed lines indicate the fields of the three-magnon resonances $2\omega_1 = \omega_\eta$. Left inset shows the dependence of the second-harmonic amplitude on the first-harmonic amplitude at $B_e = 241\,$mT, which is the resonance point for the $1+1\to5$ process. The dependence is parabolic at low oscillation amplitudes. The spatial map of magnetization oscillations at the second harmonic is shown; it corresponds to the profile of mode 5.}\label{f:2}
\end{figure}

Field dependence of the first harmonic demonstrates a weakly decreasing trend because of increasing damping rate $\Gamma_1 \sim \omega_1 \sim B_x$. A pronounced dip appears at the resonance field for the $1+1 \to 5$ confluence process (Fig.~\ref{f:2}). The dip position is slightly shifted from the three-magnon resonance field because of nonlinear frequency shift of both the interacting modes. At the same time, the amplitude of the second harmonic shows a maximum in the same field range.
 
Note that the second harmonic peak appears only if spatially-nonuniform dynamics is analyzed; we evaluate magnetization oscillations averaged in one quadrant of the disk. The total magnetization oscillations over the entire disk do not demonstrate a peak at the double excitation frequency. Thus the observation of the second harmonic is not a spurious large-amplitude signal but, instead, corresponds to another spin-wave mode at the double frequency. Plotting the  spatial profile of magnetization oscillations at the double excitation frequency (Fig.~\ref{f:2}) confirms that it is in fact mode 5. At the three-magnon resonance fields of processes $1+1 \to 3$ and $1+1 \to 4$, which are within the scale of Fig.~\ref{f:2}, we find no characteristic features in the first and second harmonic, confirming that these confluence processes are prohibited.

We plot the second-harmonic amplitude $m_z(2\omega_e)$ (representative of the final mode $\eta=5$ amplitude) as a function of the first-harmonic amplitude $m_z(\omega_e)$ in Fig.~\ref{f:2}(inset). It reveals a quadratic dependence at low modes amplitudes, as is expected from the theoretical considerations in Eq.~\eqref{e:c2-c1}. We extract the three-magnon coefficient as $V_{11,5} = 2\pi\times 0.96\,$GHz, which is reasonably close to the one calculated using Eq.~\ref{e:V112}. The discrepancy is related to the influence of other nonlinear processes on the magnetization dynamics (in particular the nonlinear frequency shift) as well as to the edge effects -- finite-difference-based micromagnetic solvers treat a curved boundary in a complicated way that is not accounted in our calculations.     

\subsection{Spatial spectrum considerations}

Another interesting point is the dependence of three-magnon coefficients on the modes indices. In bulk samples and thin films, the momentum conservation rule for degenerate three-wave confluence is $2\vec k_\nu = \vec k_\eta$. Spin-wave modes in a small-size sample have a broad spatial Fourier-spectrum instead of a single peak, but they still can be characterized by the position of the spatial spectrum maximum $\vec k_\nu$. Naturally, in the case of  a broad spatial spectrum one cannot expect a strict selection rule for $\vec k_\nu$. However, a correlation between the three-magnon coefficient and the spatial spectrum could exist. We thus calculated three-magnon coefficients for the interaction processes $\nu + \nu \to 5$, as only these processes are allowed for all $\nu$ modes. Of course, most of these processes can never be resonant due to the field dependence of their frequencies, see Fig.~\ref{f:1}. Nonetheless, nonresonant processes could also have a substantial impact on magnetodynamics, in particular, via nonlinear frequency shift \cite{Lvov_Book1994, Safonov_Book}. Our conclusions can also be applied to other samples with resonant processes.  

\begin{table}
\begin{center}
\begin{tabular}{|c|c|c|c|c|}
 \hline
  Interaction & $\vec k_\nu$ & 2$\vec k_\nu$ & $\vec k_5$ & $V_{\nu\nu,5}/2\pi$ \\
  process & ($\t{\mu m}^{-1}$) & ($\t{\mu m}^{-1}$) & ($\t{\mu m}^{-1}$) & (GHz) \\ \hline
  $1+1 \to 5$ & (0,0) & (0,0) & (86,110) & 1.58 \\ \hline
  $2+2 \to 5$ & (61,0) & (122,0) & (86,110) & 2.0 \\ \hline
  $3+3 \to 5$ & (110,0)& (220,0) & (86,110) & 0.28 \\ \hline
  $6+6 \to 5$ & (160,0) & (320,0) & (86,110) & 0.16 \\ \hline
  $4+4 \to 5$ & (0,110) & (0,220) & (86,110) & 0.91 \\ \hline
  $5+5 \to 5$ & (86,110) & (172,220) & (86,110) & 1.65 \\ \hline
\end{tabular}
\end{center}
\caption{Three-magnon interaction efficiency for the process $\nu + \nu \to 5$ in a symmetric magnetization  state. Bias field $B_x = 10\,$mT is applied along the long axis of the disk. $\vec k_\nu$ is the position of the maximum of the spatial spectrum $\hat{\mathcal F}[s_{\nu,z}]$.}\label{t:2}
\end{table}

The results summarized in Table~\ref{t:2} reveal the general trend in the relation between spatial spectrum and magnon processes. Among the $(n_{\nu,x},0)$ modes, the maximal three-magnon interaction is reached for the process $(2+2) \to 5$, which corresponds to the minimal deviation from the condition $2 k_{\nu,x} = k_{5,x}$. We find that the larger the difference $|2 k_{\nu,x} - k_{5,x}|$ is, the smaller is $V_{\nu\nu,5}$. For the $k_y$ component, such dependence is hard to point as only $n_y = 0$ and $n_y = 1$ modes are studied here. In general, the largest three-magnon interaction is expected for modes whose maximum of the spatial spectrum approaches the momentum conservation $|2 \vec k_\nu- \vec k_\eta| \to 0$.

However, it should be also pointed out that in the case of standing spin waves, the term $(\vec s_\nu \cdot \vec s_\nu)$ in Eq.~\eqref{e:V112} contains peaks not only at $2\vec k_\nu$ but also at $\vec k = 0$. This may lead to a more complex dependence of $V_{\nu\nu,\eta}$ on the mode numbers. In particular, one can expect nonvanishing interaction of a pair of high-$k$ modes with the lowest antisymmetric mode. 


\section{Symmetry-breaking perturbations}\label{s:sym_break}

\subsection{Uniform tilt of bias field}

Magnon interaction selection rules contain static magnetization configuration and spin-wave profiles. Symmetry-breaking magnetic fields applied to the sample can alter these two constituents and thus modify the magnon interaction coefficients. A uniform magnetic field, that is applied under an angle to the symmetry axis of the sample, can lead to a uniform tilt of the magnetization configuration $\vec{\mu}(\vec{r})$. For an in-plane tilt, we can assume the ``misalignment angle'' $\varphi_M$ in Eq.~\eqref{e:V-rotated} to be coordinate-independent. One finds that diagonal components $N_{xx}$ and $N_{yy}$ start to contribute to the three-magnon interaction for \emph{symmetric} final-state modes $\vec s_{\eta}$. This contribution is proportional to $\sin2\varphi_M$. For a small tilt, it therefore linearly increases with the tilt angle.

The same behavior is expected for an out-of-plane magnetization tilt at an angle $\theta_M$. In this case, the term proportional to $(N_{xx}-N_{zz})\sin2\theta_M$ appears, to which uniaxial anisotropy contributes as well. In general, a magnetization tilt also changes the symmetry of spin-wave modes. They attain a mixture of symmetric and antisymmetric components, which can affect  the  three-magnon interaction efficiency.

\subsubsection{In-plane tilt of bias field}

\begin{figure}
 \includegraphics[width=\columnwidth]{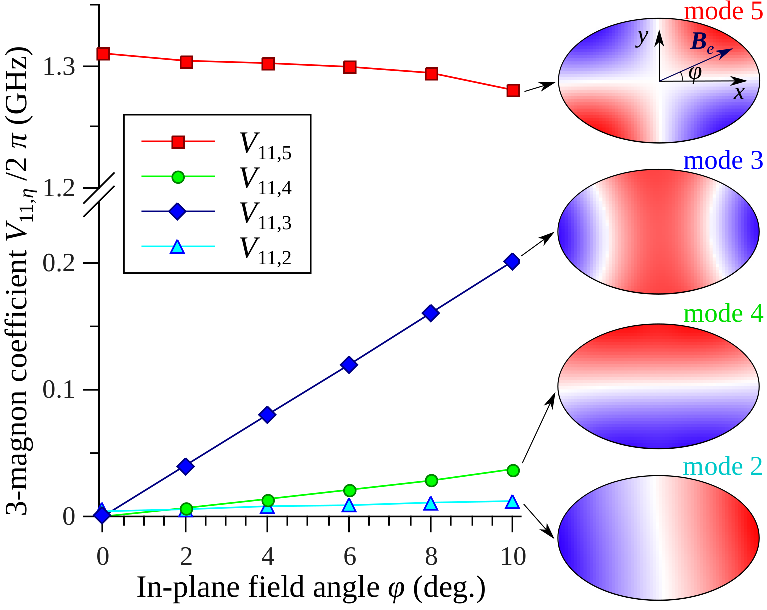}
 \caption{Dependence of three-magnon interaction efficiency $V_{11,\eta}$ on magnetic field's in-plane tilt. Note the $y$-axis break. Spin-wave profiles are shown for $\varphi = 10^\circ$.}\label{f:3}
\end{figure}

Figure \ref{f:3} shows three-magnon coefficients as a function of the field tilt $\varphi$. The coefficients are calculated at the resonance fields of their confluence processes. We find that while the tilt angle (in the presented range) does not substantially alter the mode frequencies ($\sim$50\,MHz), the three-magnon interaction is drastically affected. For all modes, that have vanishing confluence efficiency at zero field tilt, the coefficient $V_{11,\eta}$ increases linearly with the tilt angle. The strongest increase is observed for the fully symmetric mode 3. As explained above, under a field tilt, the symmetric diagonal components of the operator $\mat N$ start to contribute to three-magnon interaction, which allows for coupling to fully symmetric modes. Note that both dipolar and exchange interactions contribute to this coupling as the magnetization configuration in not perfectly uniform. 

Modes 2 and 4, which have intrinsically a mixed symmetry -- (A,S) and (S,A) respectively -- are less affected. Nonzero $V_{11,\nu}$ in their case is caused by losing the mode symmetry, which is clear from Fig.~\ref{f:3} -- the nodal lines of the modes rotate in the applied field direction (i.e. try to align parallel and perpendicular to the static magnetization direction). The efficiency of confluence into mode 5 shows a slight decrease with $\varphi$, which is explained by Eq.~\eqref{e:V-rotated}, where the leading term for this process decreases as $N_{xy} \cos^2\varphi_M$. Despite this decrease, the $1+1 \to 5$ process remains the strongest.  

\begin{figure}
 \includegraphics[width=\columnwidth]{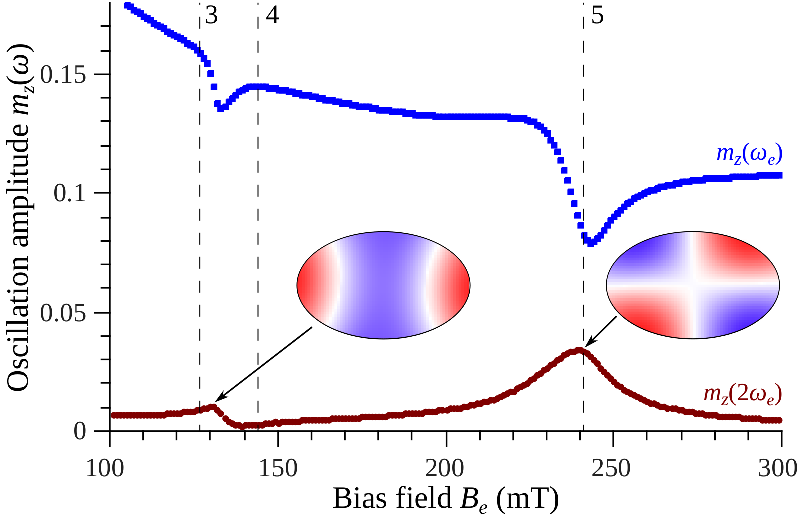}
 \caption{Amplitudes of the first and second harmonic of magnetization oscillations $m_z(t)$ excited by microwave field $b_{z} = 1\,$mT at the eigenfrequency of mode~1. The bias field is applied with an in-plane tilt of $\varphi = 10^\circ$. Vertical dashed lines indicate fields of three-magnon resonance condition. Insets show spatial distribution of the magnetization oscillations at the second harmonic.}\label{f:4}
\end{figure}

We again validate our results by inspecting the second-harmonic signal as a function of the (tilted) bias field. Figure~\ref{f:4} shows confluence process into mode 3 and into mode 5, with the characteristic dips of the first harmonic and peaks of the second harmonic. Their positions are slightly shifted from the nominal three-magnon resonance fields due to nonlinear frequency shift. From the dependence $c_3 (c_1^2)$ at $\varphi = 10^\circ$ (not shown), we extract the coefficient $V_{11,3} = 2\pi \times 0.22\,$GHz. It is very close to the value of $V_{11,3} = 2\pi \times 0.2\,$GHz calculated via Eq.~\eqref{e:V112} and shown in Fig.~\ref{f:3}. 

The low efficiency of the $(1+1) \to 4$ process does not allow for its direct observation in the second-harmonic signal -- it is overshadowed by the much more efficient $(1+1 \to 5)$ process. This peculiarity underlines the importance of  simulating the spin-wave modes profiles correctly. Even small admixture of another mode, excited far from its own resonance, could  significantly alter the calculated value of the interaction coefficient. As pointed out above, we use mode-specific spatial excitation fields in our simulations.   

\subsubsection{Out-of-plane bias field tilt}

\begin{figure}
 \includegraphics[width=0.9\columnwidth]{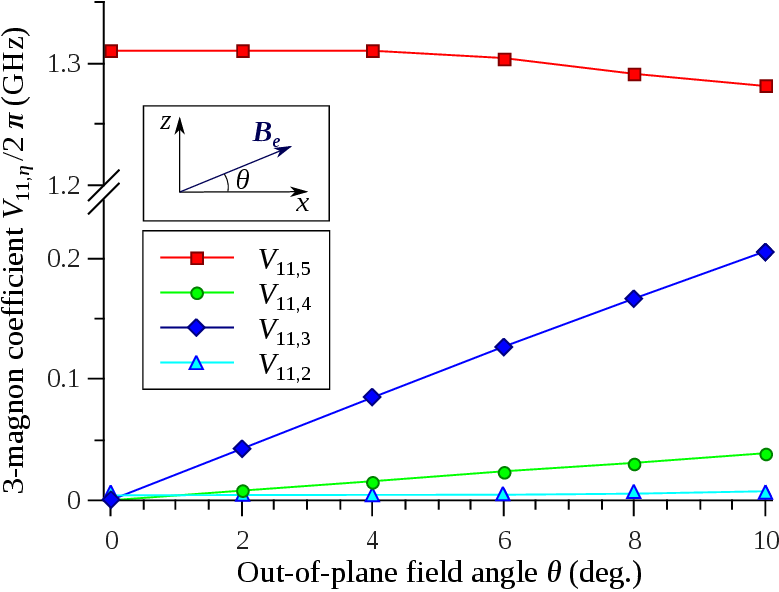}
 \caption{Dependence of three-magnon interaction coefficient $V_{11,\eta}$ on the out-of-plane tilt of the bias magnetic field.}\label{f:5}
\end{figure}

As shown in Fig.~\ref{f:5}, an out-of-plane field tilt has a very similar effect as the  in-plane field tilt. The three-magnon interaction with all modes become allowed. The process involving mode 3, which is fully symmetric in the unperturbed state, is maximally enhanced by the tilt. Nonzero values of the coefficients $V_{11,2}$ and $V_{11,4}$ are related, as in the previous case, with the breaking of the modes symmetry. This symmetry breaking is of a dipolar origin and is similar to weak nonreciprocity of spin waves in perpendicularly magnetized waveguides \cite{Wang_NatCom2021}. Although the altered mode symmetry is barely distinguishable (in plots like ones in Fig.~\ref{f:1}), it is sufficient to achieve a notable change in the three-magnon coefficient.

Comparing Figs.~\ref{f:3} and \ref{f:5}, one can point out that in-plane and out-of-plane field tilts at the same angle result in comparable values of three-magnon coefficients $V_{11,\eta}$ for $\eta = 2,3,4$. From theoretical considerations, it is clear that the tilt of the static magnetization is determinative for the three-magnon interaction, but not the tilt of the applied field. In the considered sample, in-plane field tilt causes larger magnetization tilt than the out-of-plane one. For example, at the resonance field of $1+1\to 3$ process ($|\vec B_e| \approx 128\,$mT), averaged magnetization is tilted at $\varphi_M = 7.5^\circ$ if the field applied in plane at $\varphi = 10^\circ$, and only out-of-plane angle  $\theta_M = 4.4^\circ$ is reached when the field deviates at $\theta = 10^\circ$ from the sample plane. Thus, we can conclude that in the considered case three-magnon interaction is more sensitive to out-of-plane static \emph{magnetization tilt} than to an in-plane one, which is because $|N_{zz}| > |N_{xx,yy}|$. For thin flat dots made of magnetically isotropic material this relation always holds, while presence of anisotropy, both perpendicular or in-plane, can alter this rule.

\subsection{Spatially nonuniform bias field}

In the previous subsection we considered effects of a spatially uniform tilt of the bias field. From the theoretical analysis it is apparent that application of nonuniform but spatially symmetric magnetic field, $\vec B_e(x,y) = \vec B_e(-x,y) = \vec B_e (x,-y)$, does not alter the above-formulated selection rules. Such field does not provide additional symmetry breaking of the static magnetization configuration or spin-wave modes compared to uniform field with the same components, i.e., it cannot make symmetric or antisymmetric distribution nonsymmetric. It also does not invoke any additional components of the operator $\mat N$. In this Sub-Section, we thus consider the symmetry-breaking effects of antisymmetric perturbation fields. 

\subsubsection{Gradient field}

\begin{figure}
 \includegraphics[width=\columnwidth]{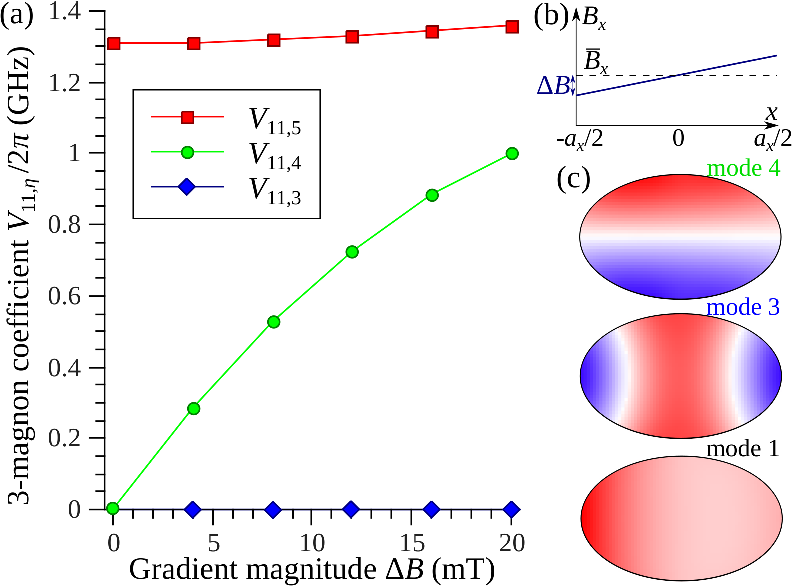}
 \caption{Effect of the spatially nonuniform field with $B_x$-gradient along $x$ axis. (a) Three-magnon interaction coefficients are shown as functions of the gradient. Bias field $B_x(x)$ varies linearly along the long axis of the disk, as shown in panel (b). The averaged field $\bar{B}_x$ is adjusted so that the three-magnon resonance condition is satisfied. (c) Mode profiles are calculated for $\Delta B = 20\,$mT.}\label{f:6}
\end{figure}

First, we apply in-plane magnetic field along the disk's major axis that has a position-dependent magnitude. As depicted in Fig.~\ref{f:6}(b), the field has a gradient along the $x$ direction, which constitutes an antisymmetric perturbation. Such perturbation does not invoke diagonal components of the operator $\mat N$ (at least when the averaged field is strong enough to maintain uniform static magnetization). The effect of the perturbation is thus limited to alteration of the spin-wave modes profiles and thereby of the three-magnon interaction.

The effect is particularly pronounced for the lowest mode 1 -- the amplitude gathers in the lower-field region of the disk (see Fig.~\ref{f:6}(c)). Other modes' symmetry along the $x$ axis is also diminished --  their profiles $\vec s_\nu$ are now neither symmetric nor antisymmetric, but contain both contributions. We thus expect that condition on the $n_x$ index of the interacting modes is relaxed. At the same time, modes' symmetry in the $y$ direction is preserved -- the requirement for the final mode being antisymmetric in the $y$ direction should remain valid. 

Calculations of three-magnon coefficients based on spin-wave profiles obtained from micromagnetic simulations confirm the expected behavior (Fig.~\ref{f:6}(a)). The processes $1+1 \to 3$ and $1+1 \to 6$ remain prohibited. Confluence into mode 4, which  has (S,A) symmetry in the absence of perturbation, now shows an enhanced efficiency $V_{11,4}$. Despite the maximum gradient of the magnetic field studied  being just $14\%$ (20~mT at $\bar B_x \approx 140\,$mT), its effect on magnon interaction is substantial, which demonstrates that this perturbation method is more efficient than tilting the field.

A field gradient along the $y$ direction, $\vec B_e = (B_x + \Delta B_x(y)) \vec e_x$, would have an analogous effect, promoting the confluence into modes of initially (A,S) symmetry -- e.g. $(1+1) \to 2$ and $(1+1) \to 6$. Other processes would remain prohibited. A more complex perturbation field with broken symmetry in both $x$ and $y$ directions, $\Delta B_x(x,y) \neq \Delta B_x(-x,y) \neq \Delta B_x(-x,-y)$, would enable all confluence processes, in particular those into intrinsically fully symmetric (S,S) modes.

\subsubsection{Nonuniform field tilt}

\begin{figure}
 \includegraphics[width=\columnwidth]{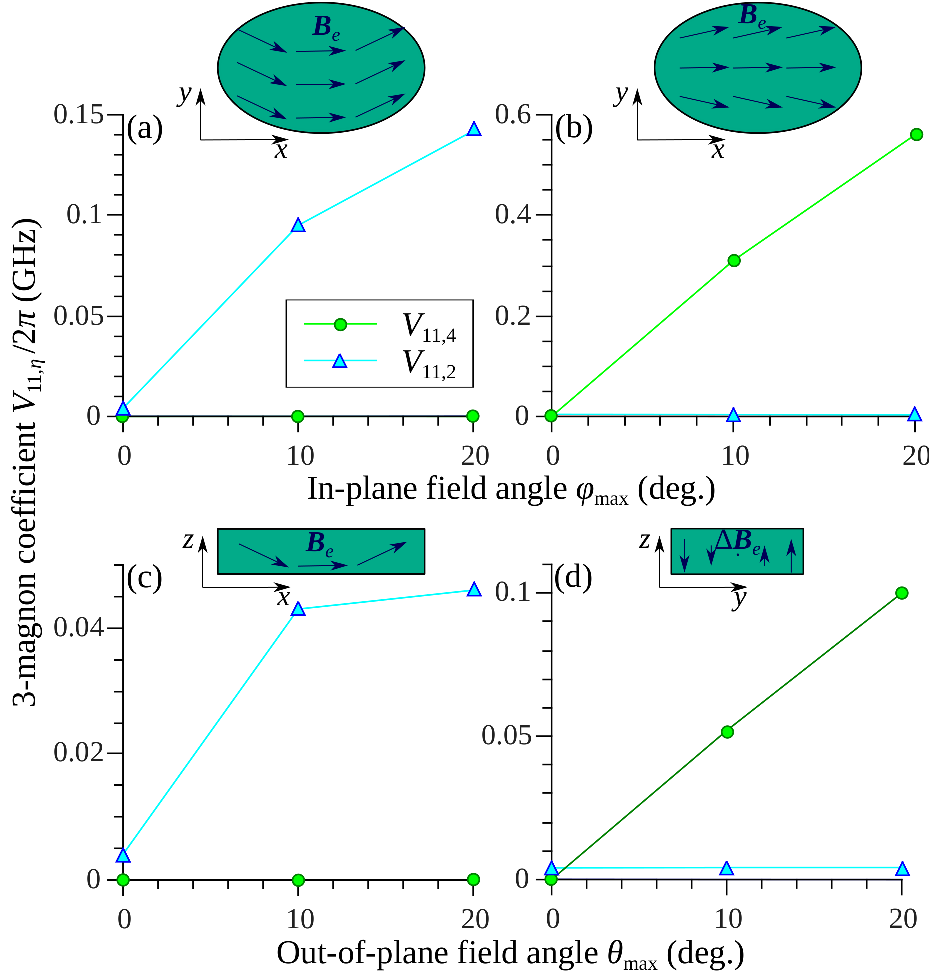}
 \caption{Effect of antisymmetric bias tilt field on three-magnon interaction coefficients. The field is $\vec B_e = B_x \vec e_x + \Delta \vec B$, where $\Delta \vec B$ varies linearly with $x$ or $y$ axis: (a) $\Delta \vec B = B_y (x) \vec e_y$, (b)  $\Delta \vec B = B_y (y) \vec e_y$, (c) $\Delta \vec B = B_z (x) \vec e_z$, (d) $\Delta \vec B = B_z (y) \vec e_z$. Field nonuniformity is characterized by the maximum tilt angle $\varphi$ (a,b) or $\theta$ (c,d) at the edge of the sample. The coefficient $V_{11,3}$ is not shown since it is not affected by the shown fields and remains negligibly small. }\label{f:7}
\end{figure}

Here, we consider a more complex but technologically-relevant \cite{Etesamirad_ACSAMI2021} symmetry-breaking field with a nonuniform tilt -- a tilt with an antisymmetric profile. We implement a uniform bias field $\vec B_e = B_x \vec e_x$ and a perturbation $B_y(\vec r)$ or $B_z(\vec r)$ with linear coordinate dependence, i.e. $B_y = B_{y,\t{max}}\cdot(2x/a_x)$ (sample center is the coordinate origin). As shown in Fig.~\ref{f:7}, such perturbation field tilt antisymmetric in the $x$ direction allows for the confluence into mode 2 and mode 6, i.e. into (A,S) modes. A $B_z(x)$ perturbation field has the same effect. Perturbation fields antisymmetric in the $y$ direction, on the other hand, enable the $1+1 \to 4$ process, i.e. confluence into (S,A) modes. Other three-magnon processes are not affected.

We parameterize the nonuniform magnetic field by the maximal angle of the field tilt $\varphi$ or $\theta$, which allows us to compare these results with the case of spatially uniform field tilt (Figs.~\ref{f:3},\ref{f:5} vs. \ref{f:7}). While uniform in-plane and out-of-plane field tilts result in comparable values of $V_{11,3}$ (for our particular geometry and material parameters), nonuniform in-plane field tilt produces a more pronounced effect than the nonuniform out-of-plane tilt.

Again, these observations can be explained by analyzing Eq.~\eqref{e:V-rotated}. When the static magnetization is tilted away from the axis of symmetry ($x$ in our case), symmetric diagonal components of the tensor $\mat N$ start to play a role. An antisymmetric profile of the magnetization tilt ($\sin \varphi_M$ or $\sin\theta_M$) is integrated with the mode profile. If the latter is an antisymmetric function, the integral returns a nonzero value.  This explains why an antisymmetric field tilt in the $x$ or $y$ direction allows for the confluence into (A,S) or (S,A) modes, respectively.

Following this argument, we also find that a field tilt antisymmetric in both $x$ and $y$ directions does not lead to an additive effect and, instead, the effects in two directions cancel each other. For such perturbation field, confluence only into (A,A) modes should be allowed, which is allowed without any perturbations anyway. We already discussed this peculiarity when considering the ``leaf state'' in Sec.~\ref{s:sym}.    

\subsection{Summary of symmetry-breaking effects}\label{ss:sym_break_summary}

The effect of symmetry-breaking perturbation fields is summarized in Table~\ref{t:3}. Antisymmetric perturbations have a mode-selective effect and allow for the confluence into modes with specific symmetry -- in addition to confluence into fully antisymmetric (A,A) modes which is always allowed. Uniform in-plane and out-of-plane field tilts are less selective -- they open confluence into mixed-symmetry modes, but to a lesser extent than into fully symmetric modes.

Table~\ref{t:3} is one the central results of this work. It can serve as a guide to find which perturbation field is required to open a particular three-magnon confluence channel or, in turn, which type of imperfections should be avoided to suppress a particular confluence process.

\textbf{Additive effects.} In general, effects of symmetry-breaking perturbations are additive. If one perturbation opens one confluence channel and the second perturbation opens another channel, then concurrent action of both perturbations will enable confluence into both channels. However, in some circumstances, e.g. at some perturbation strength, different contributions may eventually cancel each other.

An exception to the additive behavior is a perturbation field in $y$ or $z$ direction that is fully-antisymmetric $B(x,y) = -B(-x,y) = -B(x,-y)$, i.e. a combination of $B_{y,z}(x)$ and $B_{y,z}(y)$. The effect of such combination vanishes. In turn, a combination of two field components that are anti-symmetric along a single axis, that differs for these two component (i.e. $B_{\alpha}(x) = -B_{\alpha}(-x)$ and $B_{\beta}(y) = -B_{\beta}(-y)$ with $\alpha\neq\beta \in \{y,z\}$), remains additive.

\begin{table}
\begin{center}
\begin{tabular}{|c|c|c|c|c|}
 \hline
   & \multicolumn{4}{c|}{\textbf{Mode symmetry}} \\ \hline
  \textbf{Perturbation field} & (S,S) & (A,S) & (S,A) & (A,A) \\ \hline
  no perturbation & - & - & - & + \\ \hline
  $B_y$ uniform (tilt) & + & weak & weak & + \\ \hline
  $B_z$ uniform (tilt) & + & weak & weak & + \\ \hline
  $\Delta B_x(x)$ antisymmetric  & - & + & - & + \\ \hline
  $\Delta B_x(y)$ antisymmetric & - & - & + & + \\ \hline
  $\Delta B_x(x,y)$ antisymmetric & + & + & + & + \\ \hline
  $B_y(x)$ antisymmetric & - & + & - & + \\ \hline
  $B_y(y)$ antisymmetric & - & - & + & + \\ \hline
  $B_y(x,y)$ antisymmetric & - & - & - & + \\ \hline
  $B_z(x)$ antisymmetric & - & + & - & + \\ \hline
  $B_z(y)$ antisymmetric & - & - & + & + \\ \hline
  $B_z(x,y)$ antisymmetric & - & - & - & + \\ \hline
\end{tabular}
\end{center}
\caption{Effect of symmetry-breaking perturbation fields on three-magnon confluence into modes of a particular symmetry. The symmetry of the final mode is characterized in its unperturbed state. Here, ``$+$'' means that the confluence process is allowed, ``$-$'' -- prohibited process, ``weak'' -- the process is allowed but with weak efficiency. Spatially-symmetric perturbation fields have the same effect as uniform fields. ``$B(x,y)$ antisymmetric'' means that the perturbation field is antisymmetric with respect to the inversion of both $x$ and $y$ axes.}\label{t:3}
\end{table}

\textbf{Splitting and nondegenerate processes.}  As  discussed above, degenerate three-magnon splitting obeys the same rules that would now apply to the initial (splitting) mode.

Similar features are also expected for nondegenerate three-magnon scattering processes $\nu_1 + \nu_2 \to \nu_3$ and $\nu_3 \to \nu_1 + \nu_2$. As discussed above, in an unperturbed state the selection rules require both $\sum_{i=1}^3 n_{x, \nu_i}$ and $\sum_{i=1}^3 n_{y, \nu_i}$ be an odd number. Perturbations that -- according to Table~\ref{t:3} -- allow for the confluence into a mode symmetric along $x$ or $y$ direction will allow for nondegenerate scattering processes, for which the sum $\sum_{i=1}^3 n_{x, \nu_i}$ or $\sum_{i=1}^3 n_{y, \nu_i}$ is an even number, respectively.

\textbf{Other types of perturbation.} In this work, we limit ourselves to the symmetry-breaking effects of applied magnetic fields. This approach is very promising since it allows for dynamic and tunable manipulation of the magnon processes. However, symmetry-breaking effects can also be achieved by modification of sample shape (by making it less symmetric, e.g. egg-shaped) or by spatial modification of sample's magnetic parameters (such as saturation magnetization or anisotropy). For instance, in the case of fully saturated magnetization configuration, we expect the effect of magnetic parameters varying spatially along an axis to be similar to the effect of a magnetic field that breaks the symmetry of a magnon mode along the same axis. Another symmetry-breaking perturbations could be an atypical anisotropy with anisotropy axis not aligned to any of the sample symmetry axes. Finally, Dzyaloshinskii-Moriya interaction has also symmetry-breaking effect, its effect on three-magnon scattering in thin films has been discussed in \cite{Verba_PRB2019}. Concurrent application of different types of perturbations may become more complicated, beyond simple additivity, and would warrant further considerations.

\textbf{Perturbations on other magnetization states.} The above discussion also pertains to the case when bias magnetic field and static magnetization (unperturbed) are aligned parallel to the $y$ axis. The results of Table~\ref{t:3} are directly applicable to this case with the coordinate permutation $x \leftrightarrow y$. 

The case of perpendicularly magnetized unperturbed state reveals a different behavior. As discussed above, three-magnon scattering is prohibited in an ideal perpendicular state (unless an extra-ordinary anisotropy or Dzyaloshinskii-Moriya interaction are present). While detailed consideration of this case lies out of the scope of this work, some conclusions can be easily made. In particular, uniform $B_z$ tilt over $\vec\mu = \vec e_x$ state is the same as a perturbation field $B_x$ applied to the perpendicular state (note that we never invoke the smallness of perturbation in the above analysis). Thus, a uniform (and spatially symmetric) tilt of the magnetization away from the $z$ direction opens all the confluence channels (with varying efficiency). This tilt (e.g. due to stray fields of neighboring magnetic elements) could be a contributing factor to the three-magnon scattering observed experimentally in perpendicular magnetic tunnel junctions \cite{Barsukov_SciAdv2019} 

\subsection{Routes to experimental realization}\label{ss:example}

To generate a local magnetic field at the position of the sample, another small magnetic element can be placed in the vicinity. The stray fields from this auxiliary magnet can be engineered and dynamically switched/tuned to achieve a required perturbation field.

\begin{figure}
 \includegraphics[width=0.8\columnwidth]{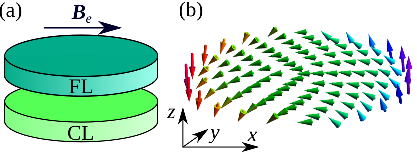}
 \caption{(a) The magnetic disk (FL=free layer) experiences the stray field (b) from the the auxiliary (CL=control layer) disk. }\label{f:8}
\end{figure}

We model a simple scenario of an auxiliary magnetic disk 
(control layer) underneath the sample disk (free layer) as shown in Fig.~\ref{f:8}(a). If the control layer is in a saturated state, its stray fields at the position of the free layer have a symmetric $B_x$ component and a fully-antisymmetric $B_y$ component -- both do not affect three-magnon interaction. However, the field also has a $B_z$ component antisymmetric in the $x$ direction only (Fig.~\ref{f:8}(b)). As detailed in Table~\ref{t:3}, we thus expect opening of confluence channels into (A,S) modes.

In our micromagnetic simulations, we choose disk separation of $1.5\,$nm and saturation magnetization of $\mu_0M_s = 1\,$T for the control layer. Our conclusions are confirmed -- we obtain substantial confluence coefficients $V_{11,2} = 2\pi \times 0.2\,$GHz for the $1+1\to 2$ process and  $V_{11,6} = 2\pi\times 0.27\,$GHz for the $1+1\to 6$ process, while confluence into modes 3 and 4 remains prohibited.

The magnetic parameters of the auxiliary layer should be engineered such as to prevent hybridization of magnetization dynamics of both layers \cite{STFMR, RodriguezPRR2022}. For that purpose, in our simulations we simply employed additional magnetic anisotropy in the $x$ direction for the control layer, which pushes control layer modes to higher frequency range. 

The state of the control layer can be, in principle, varied dynamically, thus allowing for dynamic control of three-magnon splitting and confluence. For somewhat larger control layers, this could be done by utilizing vortex-to-saturated state transition under applied field or current. The control layer can also be replaced by a nanoscale synthetic antiferromagnet. In its normal state, the stray fields are vanishingly small, whereas triggering its spin-flop transition would switch on a nonuniform stray field. Recently, this approach has been experimentally realized in Ref.\,\cite{Etesamirad_ACSAMI2021}. Other approaches involving spin-torque, heat, and voltage-controlled anisotropy for dynamic control of magnon scattering can be envisioned.


\section{Summary}\label{s:summary}

In summary, this work present a detailed theoretical/numerical study of magnon processes in laterally confined thin-film magnets with discrete magnon spectrum. The main focus lies on degenerate three-magnon confluence processes, in which two magnons fuse into one new magnon.

Our theoretical framework on the basis of the vector Hamiltonian formalism \cite{Tyberkevych_ArXiv} describes magnon interaction through an overlap integral that contains an interaction operator with contributions from  exchange, anisotropy, and magnetodipolar interactions. Other contribution, for instance, Dzyaloshinskii-Moriya interaction, can be also accounted for if needed.

We find that three-magnon processes are crucially sensitive to  the symmetry of the static magnetic configuration as well as to the profile symmetry of the participating spin-wave modes. We completed a comprehensive study for a thin elliptical disk and postulate selection rules for the magnon process.

When a disk is strongly magnetized in-plane along one of its axes, only confluence into a fully-antisymmetric modes is allowed. In such highly-symmetric magnetization state, only off-diagonal components of the magnetodipolar interaction operator, which are antisymmetric, contribute to three-magnon interaction, causing the selection rules.

For degenerate three-magnon splitting processes, this and other selection rules are valid -- with the rules now applicable for the initial splitting mode. In a general case of a nondegenerate three-magnon splitting or confluence, the selection rule for the high-symmetry case is transformed to the requirement that the sum over indices $n_x$ and the sum over $n_y$ are both odd.  

Breaking the symmetry of static magnetization configuration and/or spin-wave mode profiles results in relaxing the above-formulated selection rule. Typically, symmetry-breaking perturbations are mode-selective and enable confluence into modes of specific symmetry.

We provide guidelines for designating particular perturbation fields for opening distinct three-magnon confluence channels. These guidelines range from (i)~accurate calculations of the magnon process efficiencies using magnetization configuration and spin-wave profile, to (ii)~calculations of efficiencies based on harmonic analysis of magnetization dynamics, to (iii)~relative estimates based on the ``minimum momentum detuning'' rule. 

The results of our work can be used for analyzing and engineering a variety of scenarios. For instance, the symmetry-breaking can be naturally inherent or intentionally implemented to a sample system via imperfections/defects \cite{BarsukovAPL2015} or texture \cite{SchneiderPRB2021}, adjacent perturbations \cite{Etesamirad_ACSAMI2021,gonccalves2018oscillatory}, spatial nonuniformity of magnetic properties \cite{BarsukovPRB2012}. On the other hand, the symmetry breaking can be induced via perturbation fields with auxiliary functionalities from heat or spin-torque driven dynamics, voltage-controlled anisotropy, light-control, and others. Symmetry-breaking fields are of particular importance for applications since they can be applied and tuned dynamically. This approach, for which a proof-of-principle has recently been demonstrated experimentally \cite{Etesamirad_ACSAMI2021}, opens novel avenues for functionalizing nonlinearity in spintronic applications -- controlling nonlinear response of magnetic neurons in neuromorphic applications, improving performance of spin-torque devices, and advancing magnet-based quantum information systems.

Our concept of symmetry analysis within vector Hamiltonian formalism \cite{Tyberkevych_ArXiv} is transferable to magnon processes of higher order and other model geometries. The developed framework allows for theoretical/numerical calculations of magnon processes and for determining magnon interaction coefficients from experimental data \cite{MeckenstockJAP2006}. Nonlinear magnetization dynamics in nanomagnetic systems, which are the building blocks of modern spintronics technologies, can be a nuisance to be mastered and an opportunity to create highly functional devices. This work provides the critical theoretical basis and calls upon efforts to develop the corresponding experimental tool set.

\section*{Acknowledgements}
The work was supported by US National Science Foundation through Grant No.~ECCS-1810541, by the National Academy of Sciences of Ukraine through Project No. 23-04/13-2022, and by National Research Foundation of Ukraine through Grant No. 2020.02/0261. The research leading to these results has received funding from the Norwegian Financial Mechanism 2014-2021 no 2020/37/K/ST3/02450. JK acknowledges support from the National Science Center -- Poland, Grants No.~2021/43/I/ST3/00550. IB thanks NVIDIA Corporation for their support.



%

\end{document}